\documentclass[%
reprint,
 amsmath,amssymb,
 aps,
 prl,
]{revtex4-1}
\usepackage{graphicx}
\usepackage{dcolumn}
\usepackage{bm}

\begin{document}

\title{Non-stationary resonance dynamics of the harmonically forced pendulum}

\author{Leonid I. Manevitch}

\affiliation{%
Institute of Chemical Physics, Russian Academy of Science\\
}%
\author{Valeri V. Smirnov}

\affiliation{%
Institute of Chemical Physics, Russian Academy of Science\\
}
\author{Francesco Romeo}
\affiliation{
 Department of Structural and Geotechnical Engineering, SAPIENZA University of Rome\\
}

\begin{abstract}
The stationary and highly non-stationary resonant dynamics of the harmonically forced pendulum are described in the framework of a semi-inverse procedure combined with the Limiting Phase Trajectory concept. 
This procedure, implying only existence of slow time scale, permits to avoid any restriction on the oscillation amplitudes.
The main results relating to the dynamical bifurcation thresholds are represented in a closed form. 
The small parameter defining the separation of the time scales is naturally identified in the analytical procedure.
Considering the pendulum frequency as the control parameter we reveal two qualitative transitions. One of them corresponding to stationary instability with formation of two additional stationary states, the other, 
associated with the most intense energy drawing from the source, 
 at which the amplitude of pendulum oscillations abruptly grows. 
 Analytical predictions of both bifurcations are verified by numerical integration of original equation. 
 It is also shown that occurrence of chaotic domains may be strongly connected with the second transition.
\end{abstract}

\pacs{05.45.-a}
\pacs{63.20.Pw}

\maketitle

\section{Introduction}\label{scn:Intro}
The harmonically forced pendulum is one of the basic models of nonlinear dynamics; as such, it has numerous applications in different fields of physics and mechanics \cite{Sagdeev88, Chirikov72, Arnold06, Neishtadt05}. 
There are two main directions in the study of this model. It can be handled as an application of general mathematical perturbation theory in which the integrable conservative system is the generating model \cite{Bogoliubov61, Hale63, Nayfeh04, Chirikov72, Neishtadt75}. 
In this framework, the obtained results hold mainly in quasi-linear approach and capture only the basic features of nonlinear forced oscillators: finiteness of resonance amplitude and possibility of its abrupt change due to instability of stationary states. 
The second direction, which was predominantly developed by physicists and mechanicians, deals with analytical description of the stationary states (still in the quasi-linear approximation), their stability analysis and merely numerical study of non-stationary dynamics \cite{Nayfeh04, Manevitch2005}. 
Recently it was shown that the concept of Limiting Phase Trajectories (LPTs), introduced in \cite{Manevitch07}, allows to find an efficient analytical description of highly non-stationary resonance dynamics in which the oscillator (pendulum) draws the maximum possible (at given conditions) energy from the source (periodic field). In particular, in \cite{Smirnov&Manevitch2011}, the non-stationary dynamics of a nonlinear forced oscillator was studied in the quasi-linear approximation. The goal of this work is to remove the restrictions on the amplitude of the forced pendulum's oscillations extending the results obtained in \cite{Manevitch&Romeo2015} for two weakly coupled conservative pendula.
Towards this goal, a semi-inverse approach in combination with the LPT concept is adopted and 
the analytical description of qualitative transitions in both stationary and highly non-stationary dynamics is derived. 
Furthermore, the conditions of chaotization for different oscillation amplitudes are clarified. \\
The paper is organized as follows. At first, the governing equations and the semi-inverse asymptotic approach are introduced and the stationary regimes of oscillations are studied. Then, in the following section, the non-stationary (global) dynamical transitions are discussed and identified. The numerical validation carried out by means of Poincar\'{e} sections is eventually described followed by the concluding remarks.

\section{Governing equation and asymptotic approach}\label{scn:Stationary}

We discuss the undamped dynamics of a pendulum excited by an harmonic external field and undergoing unidirectional motion. 
Corresponding equation of motion is

\begin{equation}\label{eq:Pend_1}
\frac{d^{2} q}{d t^{2}} + \sin{q}= f \sin{\left(\omega+ s\right)t},
\end{equation}

where $q$ is the angular coordinate of the pendulum, $f$ and $\omega$ are the harmonic forcing amplitude and frequency and $s$ is the detuning parameter. 
By introducing the complex amplitude of the pendulum oscillations as

\begin{equation}\label{eq:complex}
\begin{split}
\psi =  & \frac{1}{\sqrt{2}}(\frac{1}{\sqrt{\omega}} \frac{d q}{d t}+ i \sqrt{\omega} q)  \\
q =  & \frac{-i}{\sqrt{2 \omega}}(\psi-\psi^{*}),   \quad
\frac{d q}{d t} =  \sqrt{\frac{\omega}{2}}(\psi+\psi^{*}),
\end{split}
\end{equation}

equation \eqref{eq:Pend_1} can be rewritten as follows
\begin{multline}\label{eq:Expansion_1}
 i \frac{d \psi }{d t}+\frac{\omega}{2}   \left(\psi+\psi^{*} \right)  \\ 
 +  \frac{1}{\sqrt{2\omega}} \sum_{k=0}^{\infty}  {\frac{1}{(2k+1)!} \left(\frac{1}{2 \omega} \right)^{k} \left( \psi-\psi^{*} \right)^{2k+1}}  \\
 =\frac{f}{2\sqrt{2 \omega}} \left(e^{i (\omega +s) t}-e^{-i (\omega +s) t} \right)
\end{multline}
Let's search a solution of equation \eqref{eq:Expansion_1} as $\psi= \varphi e^{i \omega t} $, where $\varphi$ is a slowly changing function of the time $t$.
%
Substituting 
into equation \eqref{eq:Expansion_1} and assuming the detuning parameter $s$ to be small enough, one can consider the function $\varphi$ as a new variable depending only on slow time $\tau=s \,t$. 

Then, multiplying equation \eqref{eq:Expansion_1} by the $e^{-i \omega t}$ and integrating with respect to the "fast" time $t$, the condition providing elimination of the resonance (secular) terms is obtained as:

\begin{equation}\label{eq:Pend2}
i s \frac{\partial \varphi}{\partial \tau}-\frac{\omega}{2} \varphi+\frac{1}{\sqrt{2 \omega}} J_{1}\left( \sqrt{\frac{2 }{\omega}} | \varphi | \right) \frac{\varphi}{|\varphi |}=\frac{f}{\sqrt{2 \omega}} e^{i \tau} ,
\end{equation} 
where $J_{1}$ is the Bessel function of the first order.

It is easy to check that the function $\varphi = \sqrt{X} e^{i \tau}$ is the solution of equation \eqref{eq:Pend2} if the frequency $\omega$ satisfies the relation

\begin{equation}
\label{eq:tmp1}
s- \frac{\omega}{2}+\frac{1}{\sqrt{2 X \omega}} J_{1}\left(\sqrt{\frac{2 X}{\omega}} \right) -\frac{f}{\sqrt{2 X \omega}}=0
\end{equation}

Taking into account the relationship $X=\frac{\omega}{2} \, Q^{2}$ between oscillation amplitude $Q$ and the amplitude of its complex representation $\sqrt{X}$ (see definition \eqref{eq:complex}), 
%
one can rewrite equation \eqref{eq:tmp1} as follows

\begin{equation}\label{eq:tmp2}
\omega^{2}+ 2 s \, \omega - \frac{1}{Q} \left(2 J_{1}(Q)-f \right)=0.
\end{equation}

\begin{figure}
\includegraphics[width=85mm]{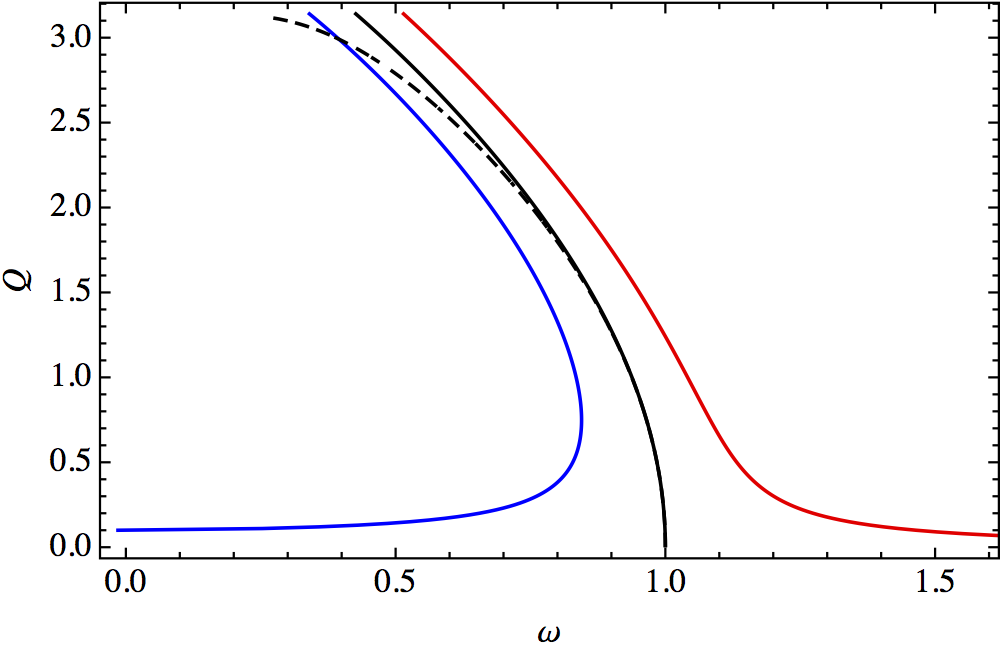}
\caption{(Color online) Frequency-amplitude curves for the pendulum stationary oscillations. 
Black, red and blue  curves show the backbone natural  and the forced frequencies, respectively.
 Detuning parameter $s = \pm 0.05$ and force amplitude $f = \pm 0.1$. 
 Solid curves correspond to equation \eqref{eq:frequency}, while the dashed one shows the exact backbone value.}
\label{fig:Frequency}
\end{figure}

This equation permits a very simple expression for the frequency of pendulum stationary oscillations as a function of their amplitude:

\begin{equation}\label{eq:frequency}
\omega=-s+\sqrt{\frac{1}{Q} \left(2 J_{1}\left(Q \right)-f \right)+s^{2}}
\end{equation}
One should note that the high-frequency branch of the forced oscillations is described by negative values  of both detuning $s$ and force amplitude $f$.
It can also be seen that the limit $\{s \rightarrow 0; \, f \rightarrow 0 \}$ leads to the backbone frequency of the natural vibrations of the pendulum $\omega_{0} = \sqrt{2 J_{1}(Q)/Q}$.
The latter approximates the exact value of natural frequency of pendulum oscillations  in a wide range of amplitudes, up to $9 \pi /10$ (see fig. \ref{fig:Frequency}).

%

Figure \ref{fig:Frequency} shows the frequency-amplitude dependence for both the forced and unforced oscillations.

Equation \eqref{eq:frequency} allows to estimate the amplitude, at which the bifurcation of stationary states occurs.
One can see from fig. \ref{fig:Frequency} that this bifurcation takes place at the extreme of low-frequency branch of the forced oscillations. So, solving the equation $d \omega /d Q =0$, the amplitude $Q$ at which the bifurcation occurs can be obtained from
\begin{equation}\label{eq:extreme}
2 Q J_{2} (Q) = f
\end{equation}
%

The expression \eqref{eq:extreme} allows to define the frequency of abrupt increasing of oscillation amplitude as the frequency of external force increases along the low-amplitude branch of equation \eqref{eq:frequency}.
However,  the abrupt decreasing of oscillation amplitude along the high-amplitude branch can not be tracked due to the absence of any limitation on the oscillation amplitude in the initial equation of motion.
As a result, the system does not prevent the motion from the transition into rotating regimes.

\section{Dynamical transitions}\label{scn:Transitions}
The results of previous section relate to states corresponding to fixed points in the system phase space.
Thus, they reflect the stationary dynamics of the forced pendulum (let's note that the term "stationary" does not always mean "stable").
However, non-stationary regimes in the dynamics of pendulum exist, which are characterized by a large modulation of the oscillations. Such processes are accompanied by the intensive energy exchange between pendulum and energy source.\\
This non-stationary dynamics of the forced pendulum can be tackled by introducing the phase $\delta(\tau)$ and the amplitude $a=\sqrt{\omega/2} Q$,  such that the slowly varying function $\varphi$ can now be expressed as $\varphi=ae^{i\delta(\tau)}$. Having introduced the  phase shift $\Delta=\tau-\delta$, the equation of motion \eqref{eq:Pend2} can be written as
 \begin{eqnarray}\label{eq:AD_1}
 & sa\dot{\Delta}+\frac{1}{\sqrt{2\omega}}J_1\left(\sqrt{\frac{2}{\omega}}a\right)-\left(s+\frac{\omega}{2}\right)a  \\ & =\frac{f}{2\sqrt{2\omega}}\cos\Delta \nonumber \\
 & s\dot{a}=\frac{f}{2\sqrt{2\omega}} \sin\Delta, 
\end{eqnarray}
where the overdot indicates the derivative with respect to the slow time $\tau$. The corresponding integral of motion reads
\begin{equation}\label{eq:Hamiltonian}
H=\frac{s}{2} 
\left[a^2 
\left(s+\frac{\omega}{2}\right)+\frac{a f \cos \Delta }
   {\sqrt{2\omega}}+J_0\left( 
   \sqrt{\frac{2}{\omega}}a
   \right)-1
   \right]
\end{equation}

\begin{figure}
\includegraphics[width=85mm]{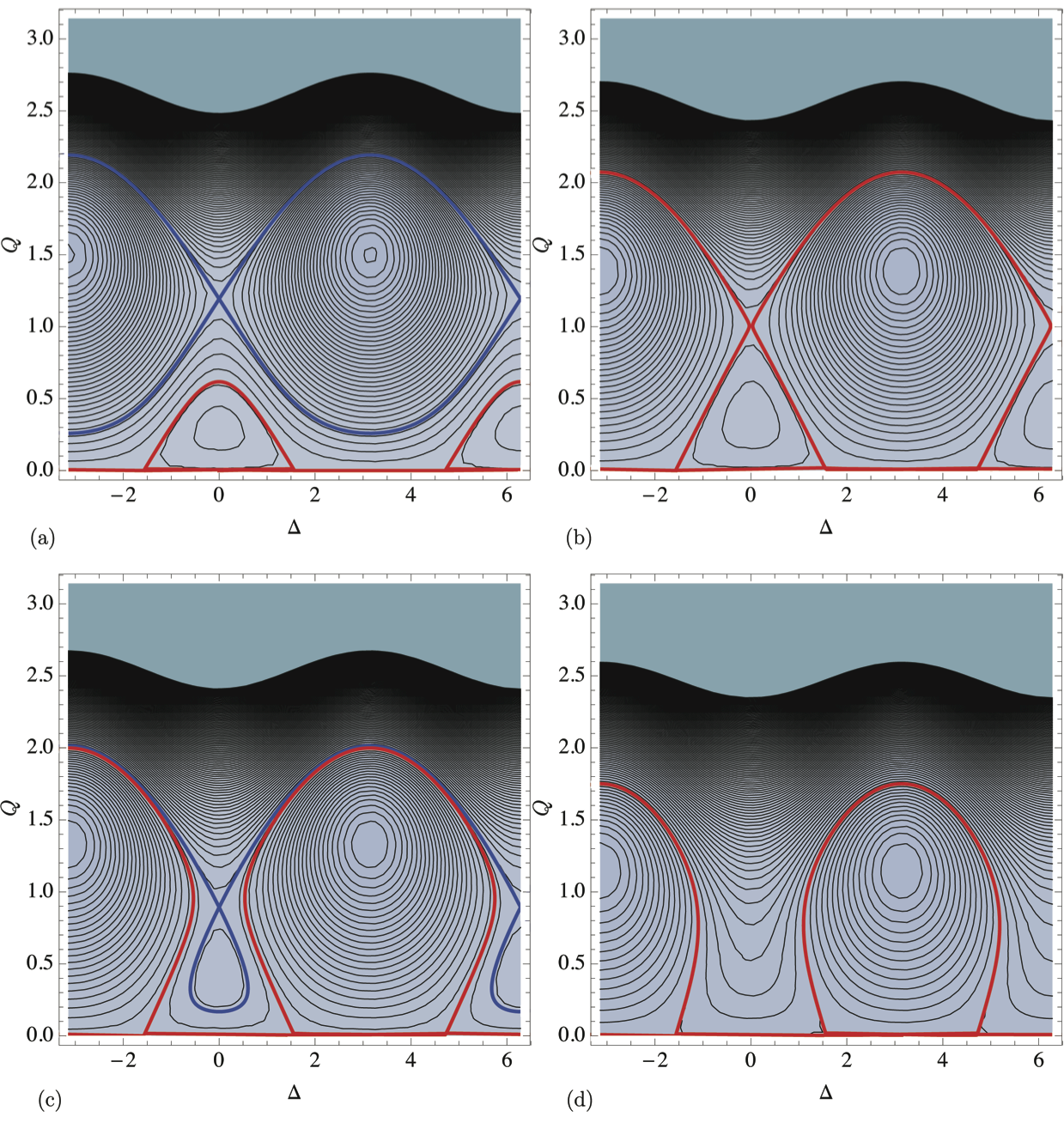}
\caption{(Color online) Evolution of the $\Delta$-$Q$ phase portrait for
$f = 0.06$ and $s=0.1$ for increasing frequency $\omega$. (a) Before the
non-stationary transition, $\omega= 0.79$; (b) at the non-stationary transition, $\omega= 0.8106$;
(c) after the non-stationary transition, $\omega= 0.82$; (d) after the stationary transition, $\omega= 0.85$.}
\label{fig:PhasePortrait1} 
\end{figure}

There are three control parameters in the considered system \eqref{eq:Hamiltonian}:  detuning $s$, forcing $f$, and frequency $\omega$. They are coupled with the stationary oscillation amplitude $Q$ by relation \eqref{eq:frequency}.
So, by fixing the detuning $s$ and the force amplitude $f$ and varying the pendulum frequency $\omega$, the phase portrait of system \eqref{eq:Hamiltonian} on the $\Delta$-$Q$ plane can be considered in order to study the non-stationary processes 
/figs \ref{fig:PhasePortrait1} (a-d)/.

\begin{figure}
\centering
\includegraphics[width=85mm]{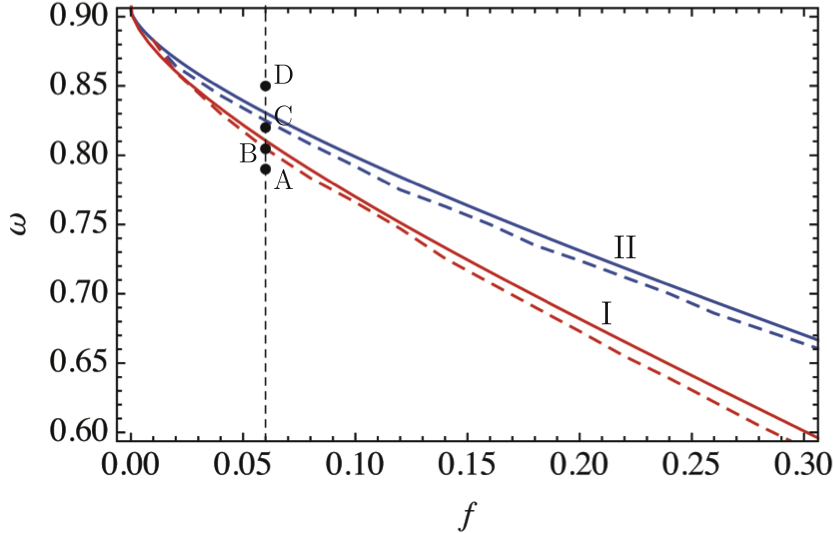}
\caption{(Color online) Dynamical transitions thresholds on the ($f$-$\omega$) plane for $s=0.1$. First (blue) and second (red) thresholds, analytical (solid), numerical (dashed), and points corresponding to the phase portraits shown in fig. \ref{fig:PhasePortrait1}.}
\label{fig:Thresholds} 
\end{figure}

Figure \ref{fig:PhasePortrait1}(a) shows the typical phase portrait for the "low" frequency region.
The three stationary points are associated with stable, i.e. the nonlinear normal modes (NNM), and unstable branches in fig. \ref{fig:Frequency}.
It is worth noticing that the attraction areas of the stable states are bounded by two specific trajectories.
The first one is the dynamic heteroclinic separatrix (blue curve) passing through the unstable state and surrounding the large-amplitude stationary point.
The second one stems form zero amplitude initial conditions (without dependence of initial phase shift) and leads to move along the closed trajectory bounding the attraction area of the stable stationary points at $\Delta=0$ /red curve in fig. \ref{fig:PhasePortrait1}(a)/. Since the latter trajectory is the farthest from the small-amplitude stationary point, we refer to it as the Limiting Phase Trajectory (LPT). The principal difference between this trajectory and the separatrix being the finite time of the corresponding period.
The LPT corresponds to the most intensive energy taking off by the pendulum from the energy source /at given initial conditions/.
The remaining trajectories require non-zero initial conditions.\\
Increasing the frequency $\omega$, the LPT grows until it coalesces with the dynamic separatrix /fig. \ref{fig:PhasePortrait1}(b)/, and an abrupt change of the LPT amplitude takes place. This qualitative transformation of the phase portrait represents the non-stationary transition, at which the LPT splits in two branches, surrounding  the two stable states with phase shift $\Delta=0$ and $\Delta=\pi$. The latter turns out to be significantly larger than the first. Therefore the amplitudes of non-stationary oscillations increases abruptly together with the energy flow from the source to the pendulum. Further increase of the frequency $\omega$ leads to to weak decreasing of LPT and separatrix, up to the annihilation of the latter occurring at the stationary transition whose frequency is determined through equation \eqref{eq:extreme}.\\
In order to estimate the threshold of the abrupt change of oscillation amplitude (i.e. the non-stationary transition), it is worth noticing that the value of Hamiltonian \eqref{eq:Hamiltonian} turns out to be equal to zero at this bifurcation point.
So, solving the equation
\begin{equation}\label{eq:first}
H(Q, \omega)|_{\Delta=0}=0
\end{equation}
jointly with equation \eqref{eq:tmp2} with respect to $\omega$ and $Q$, at fixed values $f$ and $s$, one can calculate the threshold frequency as a function of detuning $s$ and forcing $f$.
Figure \ref{fig:Thresholds} shows the threshold values of $\omega$ at detuning $s=0.1$ as a function of forcing $f$ (solid red curve).
The bifurcation value of $\omega$ that leads to the annihilation of unstable stationary point, given by  \eqref{eq:extreme}, is also represented in fig. \ref{fig:Thresholds} by solid blue curve; the corresponding dashed curves correspond to the same thresholds obtained numerically.
 
\section{Poincar\'{e} sections}\label{scn:Poincare}
\begin{figure}
\centering
\includegraphics[width=8.7 cm]{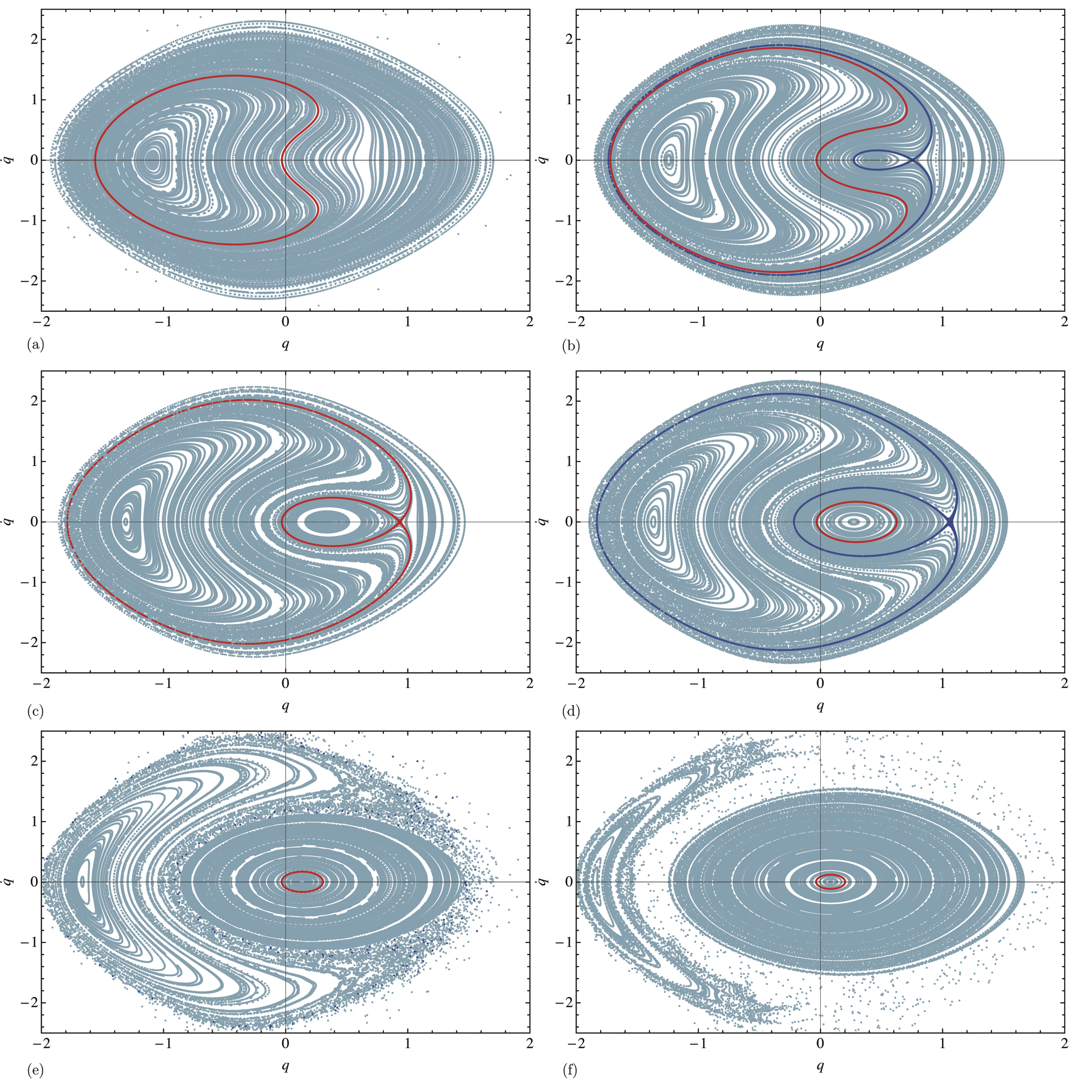}
\caption{(Color online) Evolution of the Poincar\'{e} sections for
$f = 0.06$ and $s=0.1$ for decreasing frequency $\omega$. (a) Before the
stationary transition, $\omega= 0.85$; (b) after the stationary transition, $\omega= 0.82$;
(c) at the non-stationary transition, $\omega= 0.8047$; (d) after the non-stationary transition, $\omega= 0.79$; (e) $\omega= 0.7$; (f) $\omega= 0.6$.}
\label{fig:Poincare1} 
\end{figure}

\begin{figure}
\centering
\includegraphics[width=8.7 cm]{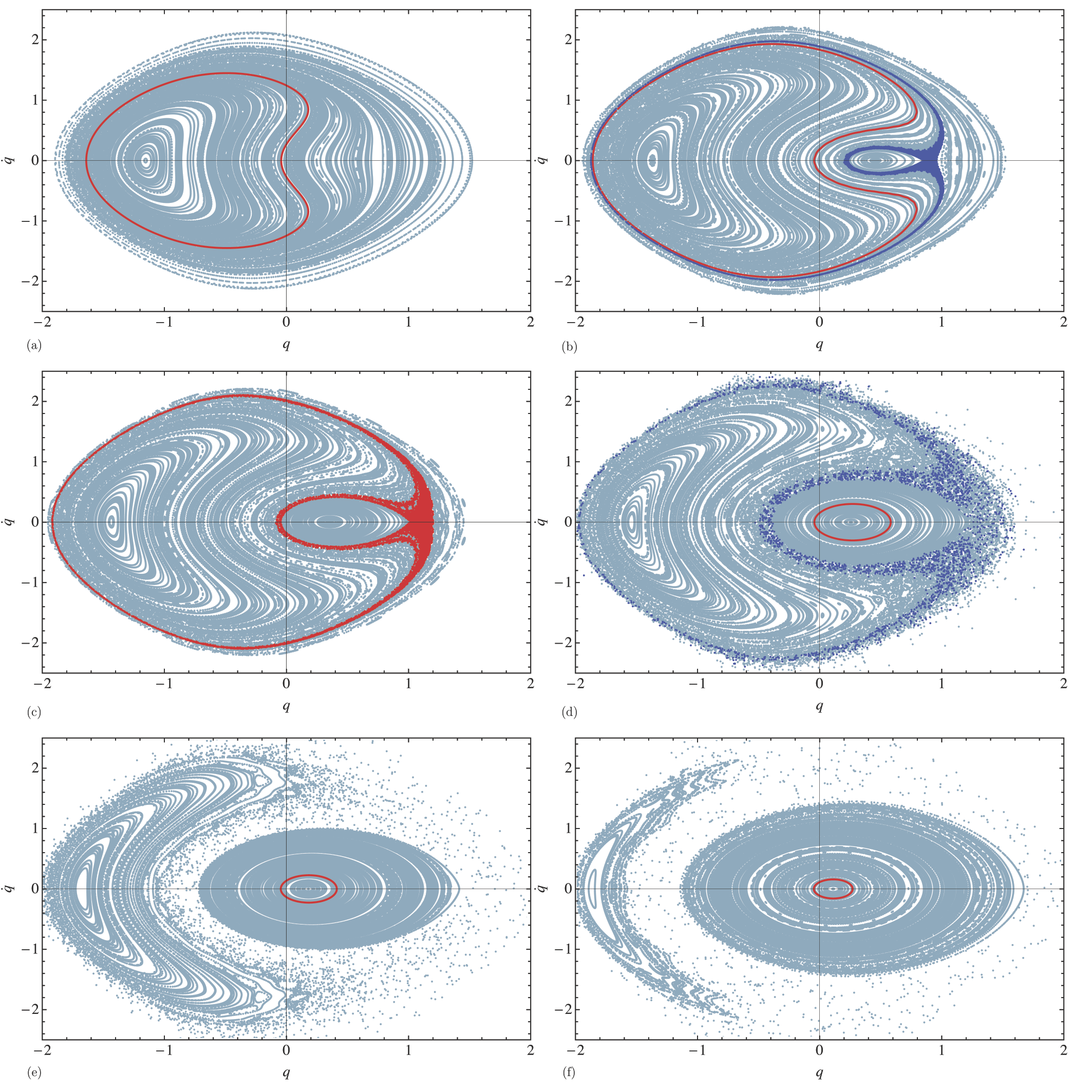}
\caption{(Color online) Evolution of the Poincar\'{e} sections for
$f = 0.08$ and $s=0.1$ for decreasing frequency $\omega$. (a) Before the
stationary transition, $\omega= 0.85$; (b) after the stationary transition, $\omega= 0.8$;
(c) at the non-stationary transition, $\omega= 0.7834$; (d) after the non-stationary transition, $\omega= 0.75$; (e) $\omega= 0.7$; (f) $\omega= 0.6$.}
\label{fig:Poincare2} 
\end{figure}

In this section a numerical validation of the dynamic regimes exhibited by the forced pendulum is proposed by resorting to Poincar\'{e} sections obtained from direct integration of the starting equation of motion \eqref{eq:Pend_1}. 
Besides confirming the two dynamical transitions discussed in the preceding section, the Poincar\'{e} sections allow to identify the onset of nonregular pendulum response and its connection with the LPTs and the dynamic separatrix. 
Having fixed the forcing amplitude and shift, the dynamic regimes evolution is the described for varying forcing frequency values. 
For $s=0.1$, in figs. \ref{fig:Poincare1} and \ref{fig:Poincare2} Poincar\'{e} sections are shown for $f=0.06$ and $f=0.08$, respectively. 
More in detail, in fig. \ref{fig:Poincare1}a the scenario at $\omega=0.85$ corresponding to point $D$ in fig. \ref{fig:Thresholds}, whose phase plane is shown in fig. \ref{fig:PhasePortrait1}d, is depicted; it is characterized by the presence of one stationary point (NNM at $\Delta=\pi$), and the LPT (red curve) encircles it.  
Next, in fig. \ref{fig:Poincare1} b, the Poincar\'{e} sections at $\omega=0.82$, corresponding to point $C$ in fig. \ref{fig:Thresholds}, whose phase plane is shown in fig. \ref{fig:PhasePortrait1} c, is shown; in this case, having crossed the second (stationary) transition occurring at $\omega=0.825$, the newborn NNM at $\Delta=0$ and unstable hyperbolic point can be seen together with the LPT (red curve) encircling the NNM at $\Delta=\pi$ and the heteroclinic separatrix (blue curve). 
As the value of $\omega$ is lowered to $0.8047$, the first (non-stationary) transition, corresponding to point $B$ in fig. \ref{fig:Thresholds} and phase-plane in fig. 3b, occurs; the LPT (red curve) and the heteroclinic separatrix coalesce implying the most intense energy exchange between the source of excitation and the pendulum. 
At last, after the second (stationary) transition the LPT  (red curve) localization is clearly seen in fig. \ref{fig:Poincare1} d in which, for $\omega=0.79$, consistently with the phase plane shown in fig. \ref{fig:PhasePortrait1}a, the LPT amplitude undergoes a significant reduction entailing a weak energy exchange with the harmonic forcing. 
For lower values of $\omega$ the regular motion region, so far characterizing the whole phase plane, splits into two regions surrounding the two stationary points, separated by a chaotic sea (see figs. \ref{fig:Poincare1}e and \ref{fig:Poincare1}f). 
The Poincar\'{e} sections reported in fig. \ref{fig:Poincare2}, corresponding to the case with $f=0.08$, show the analogous qualitative evolution of the previous case with $f=0.06$. 
The main difference lies on the onset of separatrix chaos that for $f=0.08$ occurs for higher values of $\omega$, in between the two dynamic transitions (see fig. \ref{fig:Poincare2}b). 
As shown in fig. \ref{fig:Poincare2}c, at the non-stationary transition such chaotization involves the LPT as well. 
Afterwards, as $\omega$ decreases, a trend similar to the case for $f=0.06$ can be observed.    
It is seen that for relatively weak forcing amplitude (fig. \ref{fig:Thresholds}) one can observe a regular behavior which corresponds exactly to analytical predictions while increase of forcing amplitude leads to manifestation of chaotic behavior (fig. \ref{fig:Poincare1}, \ref{fig:Poincare2}). 
Two scenarios of chaotization are possible. 
In the first scenario, this phenomenon appears after both dynamical transitions (fig. \ref{fig:Poincare1}e) and may be identified as breaking of regular separatrix. In the second, realized around $f = 0.08$ (fig. \ref{fig:Poincare2}b, c, d, e and f), chaos  manifestation occurs after the stationary transition (fig. \ref{fig:Poincare2}b) and becomes especially clear at the non-stationary transition (fig. \ref{fig:Poincare2}c).

\section{Conclusions}
By adopting a self-consistent semi-inverse procedure and LPT concept the fundamental solutions of both stationary and highly non-stationary dynamics were derived for the harmonically forced pendulum. In particular, the analytical representation of the dynamical pendulum states was found without any restrictions on the oscillations amplitude. Two qualitative transitions (in parametric space) were revealed, one of them corresponding to the birth of a pair of additional, one stable and one unstable, stationary states and heteroclinic separatrix. Manifestation of the other transition is connected with coalescence of dynamic separatrix and LPT entailing an abrupt growth of oscillation amplitude. As a result, at this purely non-stationary transition, the conditions of maximum energy that can be drawn from the source are realized. Furthermore, we have shown that chaotization of dynamic trajectories is clearly manifested in the vicinity of the non-stationary transition. The described highly non-stationary resonance dynamics of the forced pendulum lends itself to be exploited in many applications involving manipulation of mechanical energy transfer.

\acknowledgments
Authors /L.I.M and V.V.S./ are grateful the Russia Basic Research Foundation /grant n. 16-02-00400/ for the financial support.

\bibliography{F_pendulum.bib}

\end{document}